\documentclass[aps,prl,twocolumn,10pt]{revtex4-1}

\usepackage{graphicx}
\usepackage{amsmath}
\usepackage{epstopdf}
\usepackage{color}

\newcommand{\Group}{\mathcal{G}}
\newcommand{\TR}{\mathcal{T}}

\usepackage{amssymb}
\newcommand{\ff}{\pmb{\psi}}

\begin{document}

\title{GSE statistics without spin}

\author{Christopher  H. Joyner}
\author{Sebastian M\"uller}
\author{Martin Sieber}
\affiliation{School of Mathematics, University of Bristol, Bristol, BS8 ITW, United Kingdom}

\date{\today}

\begin{abstract}
Energy levels statistics following the Gaussian Symplectic Ensemble (GSE) of Random Matrix Theory have been predicted theoretically and observed numerically in numerous quantum chaotic systems. However in all these systems there has been one unifying feature: the combination of half-integer spin and time-reversal invariance. Here we provide an alternative mechanism for obtaining GSE statistics that is based on geometric symmetries of a quantum system which alleviates the need for spin. As an example, we construct a quantum graph with a particular discrete symmetry given by the quaternion group $Q8$. GSE statistics is then observed within one of its subspectra.
\end{abstract}

\maketitle

In the 1950s and 1960s Wigner and Dyson pioneered the use of random matrices in modelling the statistical properties of the energy eigenvalues belonging to complicated quantum systems \cite{Wig55,Dys62}. The techniques they developed spawned a new field of mathematics which has since become known as Random Matrix Theory (RMT) and its application has spread far and wide to many areas of Mathematics and Physics \cite{Ake11}. In particular it was later conjectured \cite{Boh84} that the high-lying quantum energy levels of classically chaotic systems are faithful to random matrix averages.

One of the cornerstones of RMT is Dyson's \emph{three-fold way} \cite{Dys62}, which groups quantum systems without geometric symmetries into three distinct types. The first occurs if time-reversal invariance is broken, for example by a magnetic field, meaning the quantum Hamiltonian $H$ is inherently complex. The remaining two appear if there is an antiunitary time-reversal operator $\TR$ which leaves $H$ invariant, i.e. $[\TR,H]=0$. They are then distinguished by either $\TR^2=1$ or $\TR^2=-1$, in which case $H$ is real symmetric or quaternion-real respectively. For chaotic systems, RMT makes predictions in all three instances by averaging over an ensemble of Hermitian matrices with the appropriate internal structure and Gaussian weighted elements. These are referred to as the Gaussian Unitary, Orthogonal and Symplectic ensembles (GUE, GOE and GSE). We note that the number of symmetry classes can be extended to ten if additional anti-commuting symmetries are present \cite{Zir96,Alt97} but this is beyond the scope of this letter.

In systems without geometrical symmetries time-reversal invariance with $\TR^2=-1$, and hence GSE statistics, can only arise if the wavefunctions have an even number of components, commonly associated with half-integer spin. For such systems GSE statistics have been predicted and/or observed numerically in examples such as quantum billiards \cite{Ber87}, maps \cite{Kep01} and quantum graphs \cite{Bol03b}, and explained using periodic-orbit theory \cite{Bol03a,Bra12}. However to date there has been no experimental observation.

For systems with geometric symmetries the situation becomes more involved. Here the Hilbert space decomposes into subspaces invariant under symmetry transformations, and the spectral statistics inside these subspaces depends both on the system's behaviour under time reversal and on the nature of the subspace. For example 3-fold rotationally invariant chaotic quantum systems display GUE statistics within certain subspectra even if they are time-reversal invariant \cite{Ley96,Kea97,Joy12a}.

What is remarkable, and has not been observed before, is that Dyson's formalism also permits GSE statistics to be obtained without any need for spin, provided the quantum system has a certain kind of discrete symmetry. More precisely, we observe GSE statistics within the subspectra of time-reversal invariant systems with $\TR^2=1$ that are associated to so-called pseudo-real irreducible representations of the corresponding symmetry group. Our specific system is a quantum graph with an internal symmetry given by the quaternion group $Q8$.

Let us first recall some important concepts regarding discrete symmetries in quantum mechanics \cite{Ell79}. If $g$  is a classical  symmetry operation in, say, position space, a corresponding
quantum mechanical operator $U(g)$ can be defined by $U(g)\psi({\boldsymbol
r})=\psi(g^{-1}({\boldsymbol r}))$, and the Hamiltonian of the symmetric
quantum system commutes with this operator.
The unitary operators $U(g)$ form a representation of the symmetry group ${\cal G}$, i.e. they satisfy $U(g) U(g') = U(g g')$ for all $g, g' \in {\cal G}$.
It can then be shown that the operators $U(g)$ have block-diagonal form in an appropriate basis of eigenfunctions of the Hamiltonian. The blocks correspond to so-called
irreducible representations (irreps) of the symmetry group. For discrete groups there are only a finite number of irreps which we label by $\alpha$ and the
corresponding block size is the dimension $s_\alpha$ of the irrep $\alpha$. The $s_\alpha$ eigenfunctions corresponding to the same block are energy-degenerate.
If we assemble them into an $s_\alpha$-dimensional vector $|\alpha,n\rangle$ then the operators $U(g)$  act as 
\begin{equation}\label{eq: irrep basis}
U(g)|\alpha,n\rangle= M^{(\alpha)}(g)^T|\alpha,n\rangle
\end{equation}
where $n$ labels different blocks belonging to the same irrep $\alpha$ and $M^{(\alpha)}(g)$ is the matrix representing $g$ in the irrep $\alpha$.
In this way the spectrum falls into subspectra associated to the different irreps.
For example, in a system with mirror symmetry the symmetry group consists of the identity $e$ and the reflection operator $r$. Then there are two one-dimensional
irreps with $M^{(\pm)}(e)=1$ and $M^{(\pm)}(r) = \pm 1$ corresponding to wavefunctions even and odd under reflection.

Like the behaviour under time reversal, all irreps may be classified into one of three types, depending on how they are related to their complex conjugate. Firstly if there does not exist a unitary matrix $S$ such that
\begin{equation}\label{irrep type}
M^{(\alpha)}(g) = S^{-1}M^{(\alpha)}(g)^*S \hspace{7pt} \forall \hspace{3pt} g \in \Group
\end{equation}
then $\alpha$ is said to be \emph{complex}. Alternatively if (\ref{irrep type}) holds and
$S=S^T$ then $\alpha$ is \emph{real} as all $M^{(\alpha)}(g)$ can simultaneously be made real by some unitary transformation. Whereas if (\ref{irrep type}) holds and $S=-S^T$ then an appropriate unitary transformation leads to a quaternion real form
consisting of $2\times2$ blocks {\scriptsize$\left(\begin{array}{cc}
a & b \\ -b^* & a^* \end{array}\right)$} with $a,b \in \mathbb{C}$, and the representation is called    \emph{pseudo-real.}    

This classification is important as it defines which spectral statistics appear within each subspace \cite{Dys62}. In particular we will argue in the following that spectral statistics in subspectra associated to pseudo-real irreps in time-reversal invariant systems with $\TR^2=1$ are appropriately modelled by the GSE. The main point is to show that the time-reversal operator $\TR$ is not the appropriate symmetry operator for the subspace, and instead one has to introduce a modified operator $\bar{\TR}$ that determines the symmetry properties of the subspace. For simplicity, we consider irreps with matrices consisting of a single quaternion-real $2\times 2$ block, and we take $\TR$ as the complex conjugation operator. In this case (\ref{irrep type}) holds with $S={\scriptsize \left(\begin{array}{cc}0 & 1 \\ -1 & 0 \end{array}\right)}$. Now, crucially, if $|\alpha,n\rangle$ satisfies (\ref{eq: irrep basis}) then the state $\TR|\alpha,n\rangle$ does not, since the $M^{(\alpha)}(g)$ are not real for all $g\in\Group$. In contrast, using (\ref{irrep type}), we see that $\bar{\TR}|\alpha,n\rangle = S\TR|\alpha,n\rangle$ does satisfy (\ref{eq: irrep basis}). Moreover $\bar{\TR}$ also commutes with $H$ because $S$ serves to exchange the components of $|\alpha,n\rangle$, while $H$ acts on each component individually. Hence $\bar\TR$ is the appropriate time-reversal operator to be considered in our subspace.

However $\bar{\TR}$ satisfies $\bar{\TR}^2 = S^2\TR^2 = -\TR^2=-1$, indicating that our subspace belongs to the symmetry class associated with the GSE. We also note that since $\bar{\TR}^2 = -1$, the states $|\alpha,n\rangle$ and $\bar\TR|\alpha,n\rangle$ are linearly independent and have the same energy, leading to Kramer's degeneracy \cite{Haa10}.

The present argument for the appearance of GSE statistics relies purely on identifying the appropriate RMT symmetry class, in the spirit of \cite{Boh84}. However one may extend semiclassical methods for systems without geometrical symmetries (see \cite{Ber85,Sie01,Mul09} and references therein) to explain why individual chaotic systems are faithful to these predictions \cite{Joy12a,Joy12b}. This builds upon earlier semiclassical work by Keating and Robbins \cite{Kea97} who incidentally predicted that subspectra associated to pseudo-real irreps show GOE behaviour. However they only investigated the so-called diagonal approximation, in which GOE is indistinguishable from GSE if one does not remove Kramer's degeneracy.

To exemplify how  pseudo-real subspaces can arise in systems without spin we turn to quantum graphs  \cite{Kot99,Gnu06}. Quantum graphs consist  of vertices $v$ connected by one-dimensional bonds $b=(v_1,v_2)$. Each bond has a specified length $L_b$ and the time-independent Schr\"odinger equation on each bond reads
\begin{equation}
H \psi(x)=-\frac{d^2}{dx^2}\psi(x)=E\psi(x),
\end{equation}
where $x$ defines the position on each bond. At the vertices the wavefunctions have to satisfy boundary conditions that make the Hamiltonian self-adjoint. For instance one can consider Neumann (or Kirchhoff) boundary conditions; these require that at each vertex
the wavefunctions of all adjacent  bonds are equal and their outward pointing derivatives sum to zero.

We have chosen to use quantum graphs since their spectral statistics have been shown to agree with the corresponding random matrix predictions in the limit of large, sufficiently well-connected graphs \cite{Gnu04,Ber06} (assuming that the bond lengths are rationally independent). Moreover, in practice numerical experiments agree well with RMT already for relatively small graphs \cite{Kot99}.

All symmetry operations on a graph may be given in terms of permutations of the vertices. A permutation $g$ is a symmetry if it leaves the connectivity and the bond lengths of the graph invariant, i.e., if for every bond  $(v_1,v_2)$ there is  a bond 
$(gv_1,gv_2)$ and it has equal length.

In order to observe GSE statistics we must choose a discrete group which admits a pseudo-real irrep, the smallest and thereby easiest to construct is the quaternion group
\begin{equation*}\label{Q8_group}
Q8:=\{\pm 1,\pm I, \pm J, \pm K\ : I^2 = J^2 =K^2 = IJK =-1\}.
\end{equation*}
Here all group elements can be written as products involving two generators, for example $I$ and $J$. One can show there are five irreps: Four real one-dimensional irreps given by $M(I)=\pm 1$, $M(J)= \pm 1,$ and a fifth two-dimensional
and pseudo-real irrep  given by the quaternion-real matrices
\begin{equation*}\label{Irrep 5}
M^{(5)}(I) = \left(\begin{array}{cc} i & 0 \\ 0 & -i \end{array}\right) \hspace{10pt} \mbox{and} \hspace{10pt} M^{(5)}(J) = \left(\begin{array}{cc} 0 & 1 \\ -1 & 0 \end{array}\right)\,.
\end{equation*}   

The aim is therefore to construct a quantum graph with this $Q8$ symmetry. Here we turn to a standard group-theoretical tool for visualising the structure of the group, known as the Cayley graph \cite{Ban09}. Cayley graphs can be constructed for any discrete group by taking the group elements as vertices and connecting them by bonds related to the generators. (Multiple Cayley graphs can exist for the same group if there are different possible choices for the generators.)

\begin{figure}[ht]
\centerline{
\includegraphics[width=0.19\paperwidth]{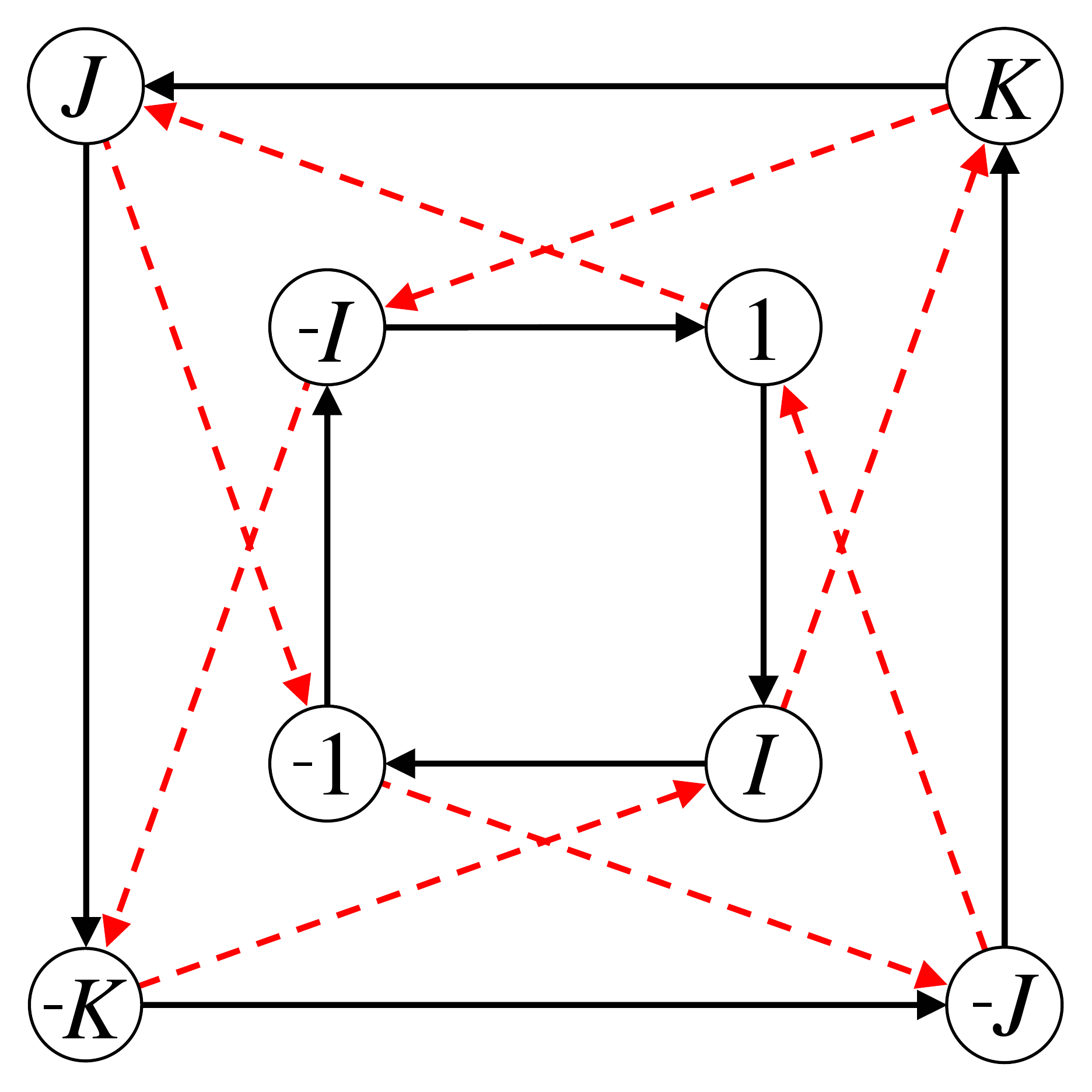}
\hspace{10pt}
\includegraphics[width=0.19\paperwidth]{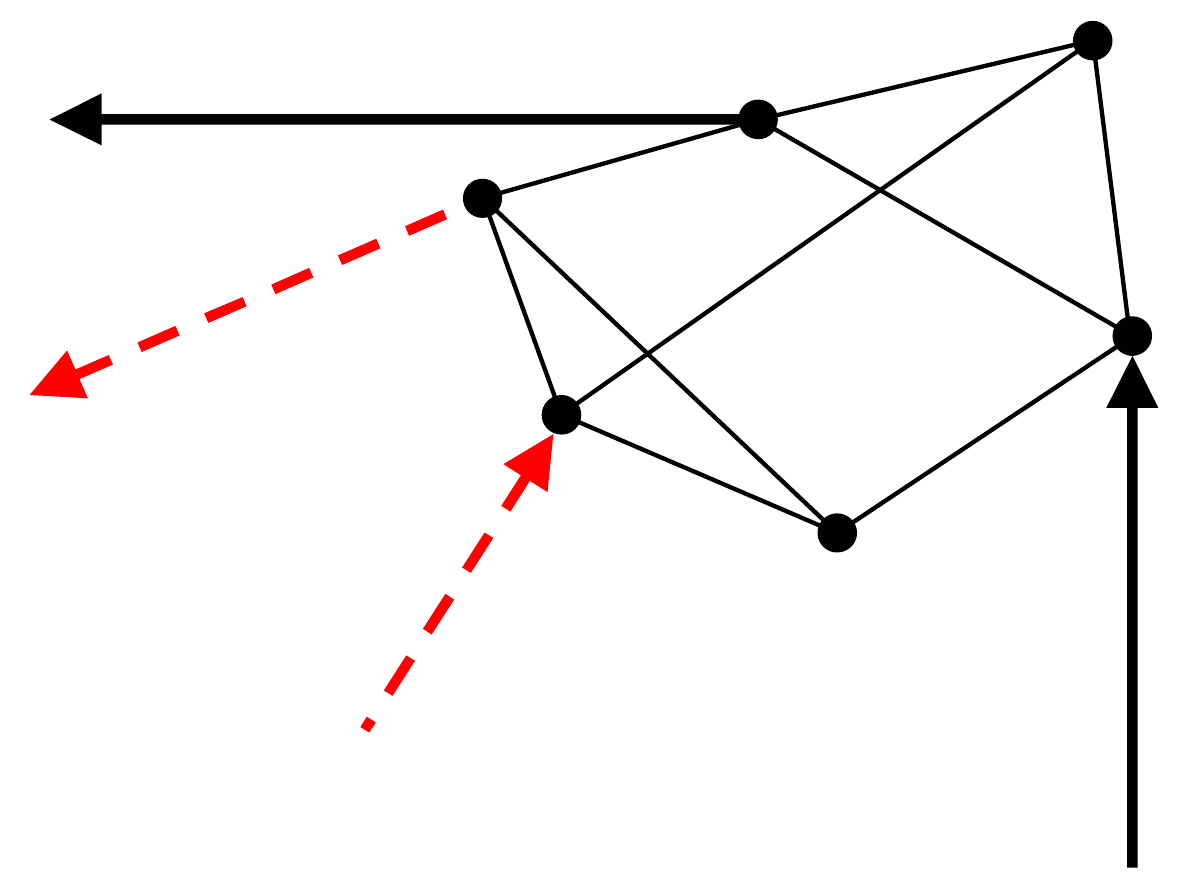}}
\caption{(a) The Cayley graph of the quaternion group $Q8$ with solid (black) bonds corresponding to the generator $I$ and dashed (red) to $J$ and (b) an example subgraph being placed at a vertex of the Cayley graph.}
\label{Cayley_Figure}
\end{figure}

In the example of Q8, see Fig. \ref{Cayley_Figure}(a), we draw bonds between two vertices representing group elements if one element can be obtained from the other by \emph{right multiplication} with either $I$ or $J$. When interpreting the result as a quantum graph we  choose the same length $L_I$ for all bonds related to $I$ and similarly $L_J$ for $J$.

The resulting graph is symmetric w.r.t. \emph{left multiplication} of all elements in $Q8$. For example, let us
consider a bond of length $L_I$ given by $b = (g,gI)$ where $g\in Q8$, then application of any element $h\in Q8$ leads to the bond $b' = (hg,hgI) = (g',g'I)$ which is also a bond in our graph with length $L_I$.  

However at present our graph is still too small to be well described by random matrix theory. A larger graph which still retains the overall symmetry can be obtained if we replace the vertices corresponding to group elements with subgraphs, see Fig. \ref{Cayley_Figure}(b). These subgraphs must be identical for each group element, meaning they must have the same number of vertices, the same connections, and the same bond lengths for analogous connections. We also need bonds connecting different subgraphs (associated to different group elements), and these have to be chosen symmetrically. To be specific let us label the vertices corresponding to the group element $g$ by $v_{g,m}$, $m\in\mathbb{N}$. Then for each subgraph $g$ at least one  vertex $v_{g,m}$ must be connected to a vertex $v_{gI,m'}$ in the subgraph $gI$. Due to symmetry the possible choices for $m$ and $m'$ and the corresponding bond lengths must be the same for all $g$. Further bonds, subject to the same symmetry conditions, have to connect subgraphs $g$ and $gJ$. 
We will discuss later that for larger subgraphs it is, in fact, advisable to connect the subgraphs by more than one bond in order to obtain a well-connected graph that displays RMT statistics.

Now, importantly, our graph has subspectra associated to each of the irreps of $Q8$. This includes the four 1D irreps of degeneracy one and the remaining 2D, pseudo-real irrep of degeneracy four (two from the dimension of the representation and two from  Kramer's degeneracy). The use of subgraphs serves to add complexity, meaning each subspectrum is expected to have RMT statistics and in particular the subspectrum associated to the pseudo-real irrep will have GSE statistics. 

The wavefunctions in our graph are scalar, alleviating any requirement for half-integer
spin.
However two-component wavefunctions (associated to the two-dimensional irrep) come into play  
if one attempts to isolate the GSE subspectrum using a so called `quotient graph'  \cite{Ban09,Par10}. This is essentially a fundamental domain equipped with the correct boundary conditions, analogous to the splitting of even and odd functions in a system with a reflection symmetry by taking the half system with either Neumann or Dirichlet boundary conditions on the former symmetry line.  

\begin{figure}[ht]
\centerline{
\includegraphics[width=0.19\paperwidth]{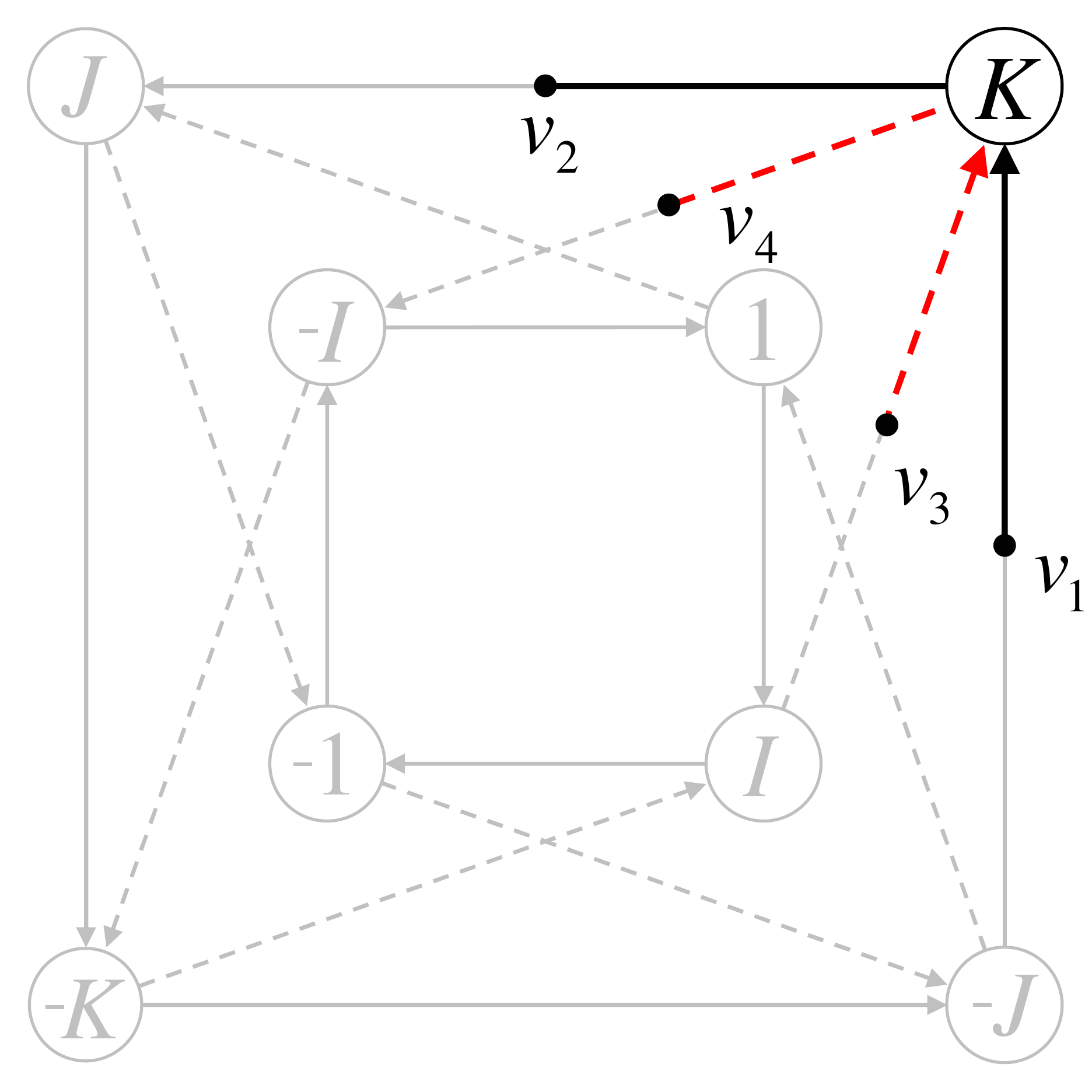}
\hspace{10pt}
\includegraphics[width=0.19\paperwidth]{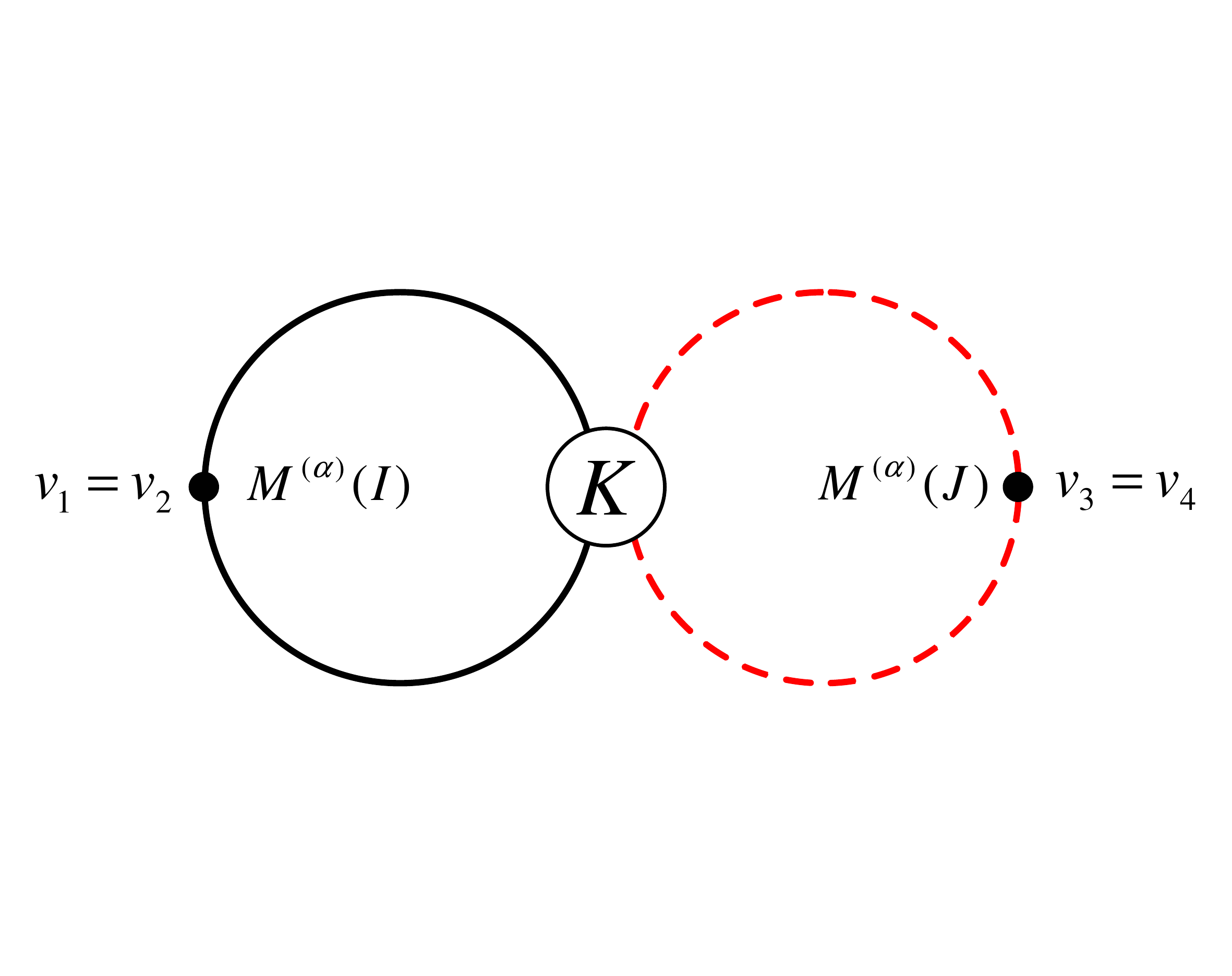}}
\caption{(a) Definition of the points $v_1,v_2,v_3,v_4$ and the fundamental domain. (b) Identification of the points $v_1$ with $v_2$ and $v_3$ with $v_4$ with vertex conditions as explained in the text leads to the quotient graph.}
\label{Quotient_Graph}
\end{figure}

We illustrate the construction of this quotient graph by starting from the Cayley graph with eight vertices, each representing one group element.
The simplest way to form the quotient graph is to take an eighth of the graph containing one vertex and half of each of the four generating bonds attached to it. We choose quite arbitrarily the vertex $K$ and cut the bonds $(-J,K)$, $(K,J)$, $(I,K)$ and $(K,-I)$ in half at the points   $v_1,v_2,v_3,v_4$ shown in Fig. \ref{Quotient_Graph} (a). These points are related by symmetry operations, in particular the application of $I$ takes the bond $(K,J)$ to $(-J,K)$ and hence the point $v_2$ in the middle of the intervening bond to $v_1$. Similarly the application of $J$ takes $(K,-I)$ to $(I,K)$ and hence $v_4$ to $v_3$. 

$v_1$, $v_2$, $v_3$, and $v_4$ now form the boundaries of the quotient graph and we have to identify boundary conditions that isolate the subspectrum associated to the pseudo-real representation. To do so it is helpful to again combine the pairs of degenerate energy eigenfunctions associated to the pseudo-real representation into vectors $\ff(x) = \langle x | \alpha,n\rangle$, where $x$ denotes a position anywhere on the graph. We now evaluate $U(I)\ff(v_1)$. The definition of $U(I)$ and the symmetries  of the graph imply $U(I)\ff(v_1)=\ff(I^{-1}v_1)=\ff(v_2)$. Combining this with $U(I)\ff(v_1)=M^{(5)}(I)^T\ff(v_1)$ (see Eq. (\ref{eq: irrep basis})) we  obtain \begin{equation}
\label{rel}
\ff(v_2)=M^{(5)}(I)^T\ff(v_1)\,.
\end{equation}
A similar result holds for the first derivatives if we let the coordinates along the bonds increase in the directions indicated by arrows in Fig. \ref{Quotient_Graph}.
In this case we obtain a relation as in (\ref{rel}) also for points moved   compared to $v_1$ and $v_2$ by the same amount, and differentiating w.r.t. this amount yields
\begin{equation}
\ff'(v_2)=M^{(5)}(I)^T\ff'(v_1).
\end{equation}
Analogous reasoning for the points $v_3$ and $v_4$ gives the conditions
\begin{eqnarray}
\ff(v_4)&=&M^{(5)}(J)^T\ff(v_3)\\
\ff'(v_4)&=&M^{(5)}(J)^T\ff'(v_3)
\label{relfinal}
\end{eqnarray}
Hence we identify $v_1$ with $v_2$ and $v_3$ with $v_4$ up to multiplication of the $\ff$ with a matrix, see Fig.  \ref{Quotient_Graph}(b).
   The relations (\ref{rel}) to (\ref{relfinal})  have the effect of isolating the pseudo-real representation,
now with two-component eigenfunctions    supported on an eighth of the original graph. The same ideas can also be applied if, as described above, the vertices of the Cayley graph are replaced by subgraphs to increase complexity and hence generate random matrix statistics. In this case the quotient graph displays GSE statistics.
Similar 'quotient systems' can be constructed to isolate the subspectra of any system with discrete symmetries. 

The conditions (\ref{rel}) to (\ref{relfinal}) provide an additional, intuitive argument for the agreement with the GSE: A wave propagating through the graph will pick up factors associated to $I$ and $J$ at the vertices $v_1$, $v_2$, $v_3$, and $v_4$. Repeated traversals of these vertices allow to sample the whole $Q8$ group, a discrete subgroup of $SU(2)$ which describes spin. This is similar to a discrete spin mechanism which, as was first noted in the context of the Dirac operator in \cite{Bol03a}, is capable of generating GSE statistics.

\begin{figure}[ht]
\centerline{
\includegraphics[width=0.15\paperwidth]{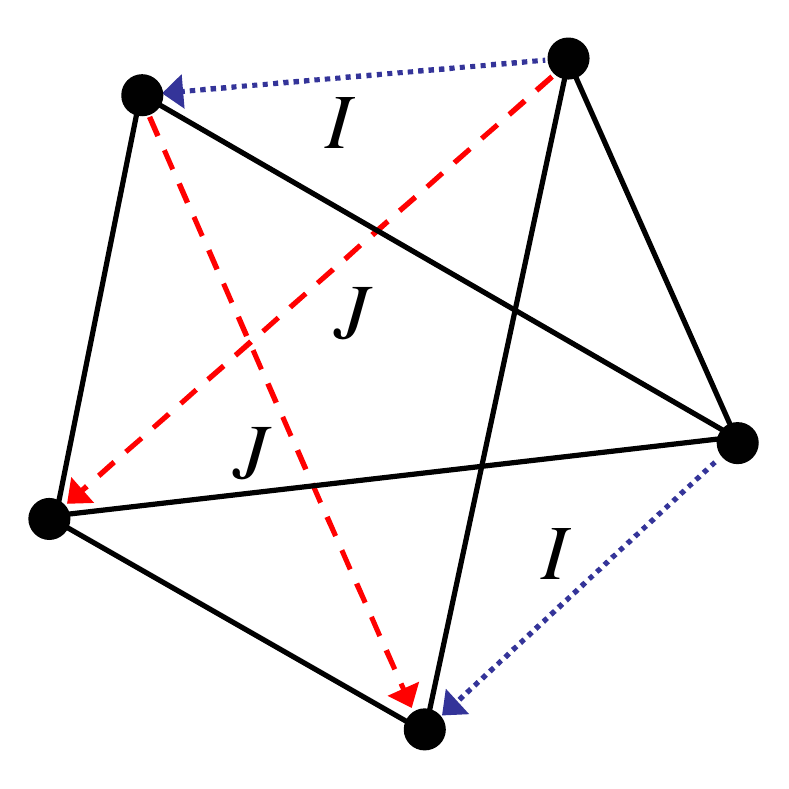}
\includegraphics[width=0.22\paperwidth]{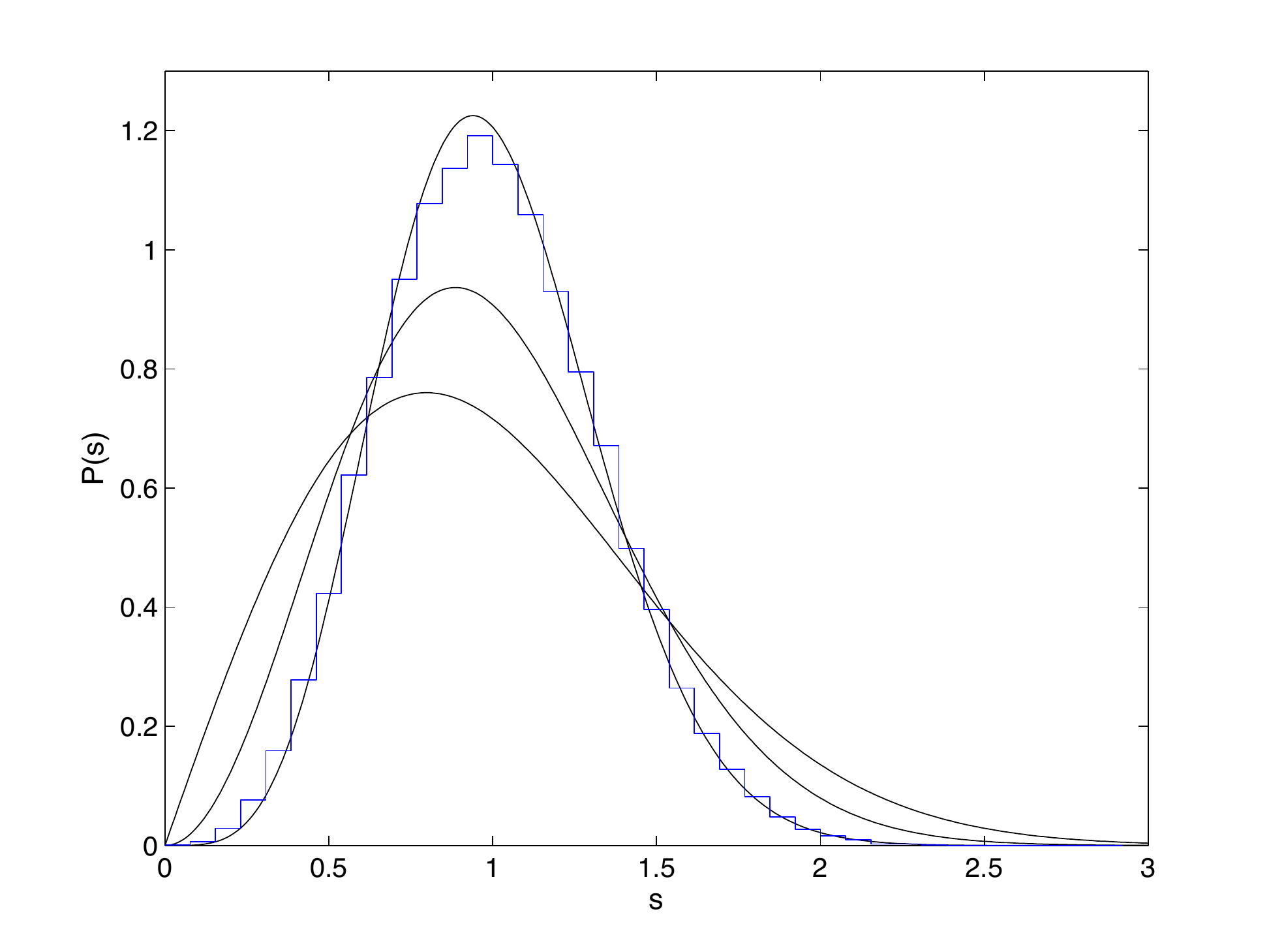}}
\caption{(a) Quotient graph containing two bonds with $M(I)$ conditions and two with $M(J)$ and (b) Nearest neighbour spacing distribution, averaged over 10 graphs with 10,000 energy levels each.}
\label{Spacing Distribution}
\end{figure}

For a numerical check we calculated the spectrum associated to the pseudo-real representation in the quotient graph displayed in Fig.\ \ref{Spacing Distribution} (a). We took an average over ten random realisations with bond lengths distributed uniformly between 0 and 1. Fig.\ \ref{Spacing Distribution} (b) shows a good agreement with Wigner's GSE prediction \cite{Wig55,Haa10} for the distribution $P(s)$ of spacings $s$ between neighbouring energy levels (normalized to yield an average spacing of 1). The distribution of each individual realisation differs only slightly from the mean. The choice of two bonds with $I$ and $J$ conditions corresponds to a better connectivity in the full $Q8$-symmetric graph (as mentioned earlier) than with only one bond. We also investigated larger graphs and found that we obtain better agreement with RMT
if they are sufficiently well-connected. This is similar as in the case of non-symmetric graphs.

In summary, we gave theoretical arguments and provided numerical evidence that GSE statistics can be observed in systems without spin if they have certain symmetry groups that allow for pseudo-real representations, as for example the quaternion group $Q8$. 
Quantum graphs as proposed here and e.g. built using optical fibres or coaxial cables \cite{Hul04} could lead to a first experimental observation of GSE statistics, avoiding the requirement of half-integer spin. It would be interesting to identify other experimental realisations of symmetries with pseudo-real representations and also investigate geometrical symmetries within the framework of the new symmetry classes \cite{Zir96,Alt97}.

The authors would like to thank R. Band, J. Harrison and U. Smilansky for helpful advice.

\bibliography{my_refs2}

\end{document}